\documentclass{appolb}
\usepackage{graphicx}
\usepackage{amsmath}
\usepackage{mathrsfs} %fuer Langrangedichte Skript - L
\usepackage{amsfonts}
\usepackage{dsfont}
\usepackage{feynmp}

\begin{document}
\title{\protect\(\tau\protect\) Vector Spectral Function from a Chirally Invariant Hadronic Model with Weak Interaction %
\thanks{ EEF70 Workshop on Unquenched Hadron Spectroscopy: Non-Perturbative Models and Methods of QCD vs. Experiment}%
}
\author{Anja Habersetzer
\address{Institut f\"ur Theoretische Physik, Goethe-Universit\"at Frankfurt}
\\
{
}
}
\maketitle
\begin{abstract}
The linear sigma model we present here describes the vacuum phenomenology of the \(N_F=2\) scalar, pseudoscalar, vector and axial-vector mesons at energies \(\simeq 1\text{ GeV}\). Together with a local \(SU(2)_L\times U(1)_Y\) symmetry obtained from a local transformation of the fields we obtain a gauge invariant effective description of electroweak interaction with hadrons in the vacuum. We show that the vector channel is described well and that the contributions from the direct decay of \(W^- \rightarrow \pi^- \pi^0\) are, although quantitatively small, necessary to reproduce the line shape of the vector spectral function. \end{abstract}
  
\section{Introduction}
The weak spectral functions of the \(\tau\) lepton as e.g. measured by the ALEPH collaboration \cite{Schael:2005am} provide an excellent ground for studying electroweak interactions at low energies. Vector meson dominance is a phenomenological approach, that attempts to describe the interaction between the photon and hadrons as \(\bar{q} q\) vacuum fluctuations of the photon. We can also understand the interaction of the charged weak bosons to hadrons in terms of such vacuum fluctuations that manifest themselves in the form of a mixing with vector and axial-vector mesons. The question whether the axial vector resonance \(a_1(1260)\) can indeed be described as a \(\bar{q}q\) state within the quark model, where it is considered to be the \(J^{PC}=1^{++}\) member of the \(N_F=2\) axial-vector multiplet, or whether it might not actually be a dynamically generated \(\rho-\pi\) molecule-like state has not been unambiguously answered \cite{Wagner:2008gz,Roca:2006tr}. Including weak interaction is thus not only useful to describe electroweak interactions of hadrons. Moreover, it also provides a basis for understanding more about the nature of the \(a_1(1260)\) resonance.

\section{Lagrangian}
Effective models as the extended Linear Sigma Model (eL\(\sigma\)M) that apply the principles of QCD to hadronic observables are widely used to study the low-energy phenomenology in heavy ion collisions. In the linear sigma model we present here scalar, pseudoscalar, vector and axial-vector fields are considered genuine degrees of freedom that can be constructed from a \(\bar{q}q\) picture. They are described by the matrix-valued fields 
\begin{align}
\Phi = (S_a+i P_a) t_a\ \!, \ L^\mu = (V_a^\mu + A^\mu_a) t_a \ \!, \ R^\mu_a = (V_a^\mu - A^\mu_a) t_a
\end{align}
on the basis of the strong isospin algebra with generators \(t_a=(1/2)\sigma_a\). 
Their masses and hadronic interactions, as well as their interaction with an electro-weak field, are obtained from the Lagrangian
{\small \begin{align}
\mathscr{L} &= \text{Tr} \bigl[(D^\mu \Phi)^\dagger (D^\mu \Phi)\bigr] - m_0^2 \text{Tr}[\Phi^\dagger \Phi]  - \lambda_1\text{Tr} [(\Phi^\dagger \Phi)]^2 - \lambda_2 \text{Tr}[(\Phi^\dagger \Phi)^2]  \nonumber \\[3pt]
& - \frac{1}{4}\text{Tr}[(L^{\mu\nu})^2 + (R^{\mu\nu})^2] + \frac{m_1}{2}\text{Tr}[\left(L^{\mu\ \!\!2}\! + \! R^{\mu\ \!\!2}\right)] +\text{Tr}[H(\Phi + \Phi^\dagger)] \nonumber \\
&+ c_1 (\text{det}\Phi - \text{det}\Phi^\dagger)^2 + i \frac{g_2}{2} \bigl( \text{Tr} \bigl[ L_{\mu \nu} [ L_{\mu},L_{\nu} ] \bigr]   + \text{Tr} \bigl[R_{\mu \nu}[ R_{\mu},R_{\nu}] \bigr] \bigr) \nonumber\\
&  + 2 h_3 \text{Tr}[\Phi R_\mu \Phi^\dagger L^\mu] + \frac{g \delta_w}{2} \text{Tr}[W_{\mu\nu}L^{\mu\nu}]+ \frac{g' \delta_\text{em}}{2} \text{Tr}[R_{\mu\nu}B^{\mu\nu}] \nonumber \\
& +\frac{1}{4} \text{Tr}[(W^{\mu\nu})^2 +(  B^{\mu\nu})^2] +\frac{g}{2\sqrt{2}}\left( W_\mu^-\bar{u}_{\nu_\tau} \gamma_\mu (1-\gamma_5) u_\tau+\text{h.c.}\right)  \nonumber\\
& + \mathscr{L}_{\Phi L R}(\Phi,R^\mu,L^\mu) + \mathscr{L}_{L R}(R^\mu,L^\mu) \ \!. \label{eq:FullLagrangian}
\end{align}}
The Lagrangian is invariant under a {\bf global} linear chiral \(U(2)_L\times U(2)_R\) transformation of the hadronic fields 
\begin{align}
\Phi \rightarrow & U_L \Phi U_R^\dagger\ \!,\ L^\mu \rightarrow U_L L^\mu U_L^\dagger\ \!, \ R^\mu \rightarrow U_R R^\mu U_R^\dagger \ \!,
\end{align}
and it is also invariant under a {\bf local} \(SU(2)_L\times U(1)_Y\) transformation by
\begin{align}
\Phi \rightarrow & U_L \Phi U_Y^\dagger \ \!,\ L^\mu \rightarrow U_L L^\mu U_L^\dagger\ \!, \ R^\mu \rightarrow U_Y R^\mu U_Y^\dagger.
\end{align}
The bare gauge fields \(W^\mu = W_i^\mu t_i\) and \(B^\mu = B^\mu t_3\) transform in the adjoint representation of \(SU(2)_L\times U(1)_Y\) as
\begin{align}
W^\mu \ &\xrightarrow{SU(2)_L} \  U_L W^\mu U_L^\dagger + \frac{i}{g} U_L \partial^\mu U_L^\dagger \ \!,\\
B^\mu \ &\xrightarrow{\ U(1)_Y\ \!} \  U_Y B^\mu U_Y^\dagger + \frac{i}{g'} U_Y \partial^\mu U_Y^\dagger \ .
\end{align}
After rotating the bare fields into the physical fields \(A_\mu\) and \(Z_\mu\) by the Weinberg mixing angle \(\theta_W\), we obtain the covariant derivative 
\begin{align}
 D^\mu \Phi  = & \ \partial^\mu \Phi - i g_1 (L^\mu \Phi - \Phi R^\mu) - i g \cos{\theta_C} (W^\mu_1 t_1 + W^\mu_2 t_2) \Phi - i e[A^\mu t_3,\Phi] \nonumber \\
& - i g \cos{\theta_W} (Z^\mu t_3 \Phi + \tan{\theta_W}\Phi Z^\mu t_3)\ .
\end{align}
The global symmetry is broken spontaneously by the scalar condensate and explicitly by the term \(\sim H\) which modulates non-vanishing but equal quark masses and also by the 't Hooft determinant that corresponds to the explicitly broken \(U(1)_A\) symmetry.
In (\ref{eq:FullLagrangian}) we included two terms \(\sim \delta_w\) and \(\sim \delta_{em}\). These terms describe a mixing between electroweak interaction fields and (axial-)vector fields. This mixing arises from the hadronic loop contributions of the electroweak bosons and is also described by the Vector Meson Dominance Model. More details on this Linear Sigma Model and its precursors on whose principles it has been developed (such as e.g \cite{Gasiorowicz:1969kn}) can be found in \cite{Parganlija:2012fy,Parganlija:2010fz,Janowski:2014ppa,Gallas:2009qp} and references therein.

\section{\protect\(\tau\protect\) and spectral functions}
\begin{figure}[h]
%\begin{fmffile}{EEF70AnjaHabersetzerMP}	\vspace{0.3cm}\hspace{0.5cm}	\parbox{60mm}{  \begin{fmfgraph*}(115,50)	    \fmfv{label=\text{\(\tau^-\)},label.dist=1}{i11}		\fmfv{label=\text{\(\nu_\tau\)},label.dist=2}{b6}      \fmfv{label=\text{\(\pi^-(k_2)\)},label.dist=1}{o2}      \fmfv{label=\text{\(\pi^0(k_1)\)},label.dist=1}{o4}      \fmfleftn{i}{13}\fmfbottomn{b}{9}\fmfrightn{o}{5}      \fmf{fermion,tension=1.5}{i11,v1}      \fmf{fermion,tension=0.8}{v1,b6}			\fmfdot{v1}			\fmf{boson,label=\text{\(W^-\)},tension=0.6,label.dist=-16}{v1,v2}			    \fmf{dashes,tension=0.45}{v2,o4}    \fmf{dashes,tension=0.45}{o2,v2}        \fmfdot{v2}  \end{fmfgraph*}} 	 	\parbox{55mm}{  \begin{fmfgraph*}(120,50)	    \fmfv{label=\text{\(\tau^-\)},label.dist=1}{i11}			\fmfv{label=\text{\(\nu_\tau\)},label.dist=2}{b6}      \fmfv{label=\text{\(\pi^-(k_2)\)},label.dist=1}{o2}      \fmfv{label=\text{\(\pi^0(k_1)\)},label.dist=1}{o4}      \fmfleftn{i}{13}\fmfbottomn{b}{9}\fmfrightn{o}{5}      \fmf{fermion,tension=1.5}{i11,v1}      \fmf{fermion,tension=0.8}{v1,b6}			\fmfdot{v1}\fmfdot{v3}			\fmf{boson,label=\text{\(W^-\)},tension=0.6,label.dist=-16}{v1,v2}			\fmf{dbl_plain,label=\text{\(\rho^-\)},label.dist=-18,tension=0.8}{v2,v3}			    \fmf{dashes,tension=0.45}{v3,o4}    \fmf{dashes,tension=0.45}{o2,v3}        \fmfdot{v2}  \end{fmfgraph*}} \end{fmffile}
 ~ \includegraphics[width=0.4\textwidth]{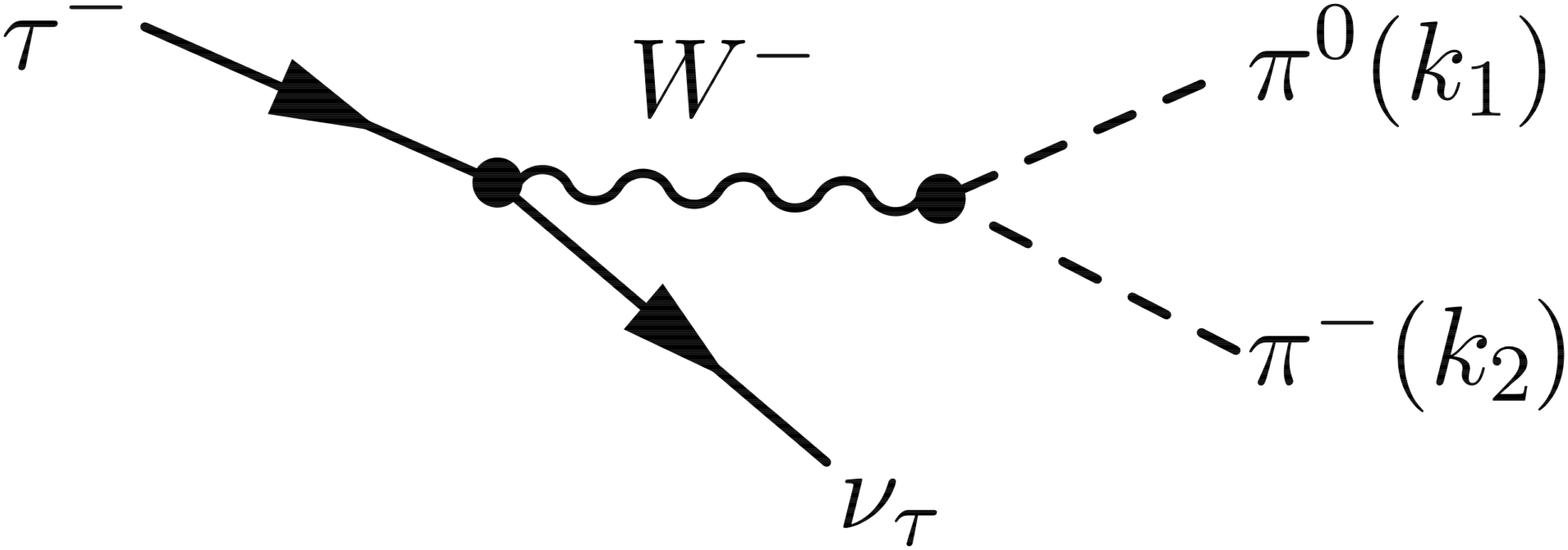} ~ ~ ~ \includegraphics[width=0.45\textwidth]{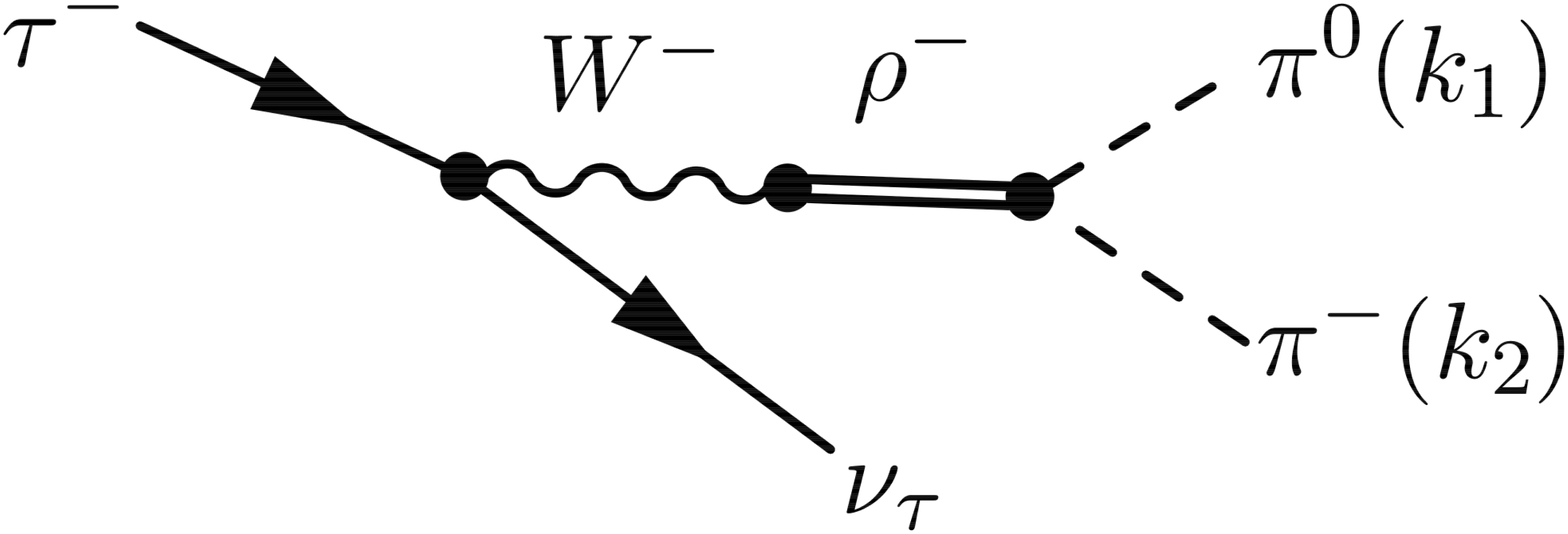}\caption{Feynman diagrams of the \(2 \pi\) decay channels of the \(\tau^-\) in the vector channel.}
\label{fig:FeynmanDiagramsVectorChannel} \end{figure}

In the vector channel the \(\tau^-\) decays dominantly into \(\pi^-\pi^0 \nu_\tau\). The contributing Feynman diagrams are shown in Fig. \ref{fig:FeynmanDiagramsVectorChannel}. The \(W\!-\!\rho\) mixing comes only from the term proportional to the new parameter \(\delta_w\). This parameter can be determined by fitting the theoretical curve to the height of the peak of the experimental mass distribution. 
Decay rates and spectral functions are calculated on tree level. We use the optical theorem \cite{Giacosa:2007bn}
that relates the tree level amplitude to the imaginary part of the self-energy \(\text{Im}[\Sigma(s)] = \sqrt{s} \Gamma (s)\) and define the vector propagator as
\begin{align}
D^{\mu\nu}(s) = \frac{-g_{\mu\nu}+\frac{P_\mu P_\nu}{P\cdot P}}{(s-m_\rho^2)^2+(\sqrt{s}\ \! \Gamma_{\rho\rightarrow \pi^-\pi^0}(s))^2} \ \!.
\end{align} 
The bare mass of \(\rho\) is given by \(m_\rho^2 = m_1^2 + \phi^2(h_1 + h_2 + h_3)/2\) and is fixed to the physical \(\rho\) mass. Therefore we request that the real part of the self-energy \(\text{Re}\Sigma(s=m_\rho^2)=0\). Then the contributions that arise from the real part of the self-energies will not contribute significantly to the vector spectral function in the region around the peak.
The sum rule
\begin{align}
\int\limits_0^\infty ds \ \! d_{\rho^- \rightarrow \pi^-\pi^0}(s) = 1 \label{eq:sumrule}
\end{align}
describes a probability conservation where \(d_{\rho}(s)\) is the probability to find the \(\rho\) meson at a given energy. The sum rule in (\ref{eq:sumrule}) holds only at all orders in perturbation theory for \(\Pi(s)\) but not necessarily if the continuum contributions to the mass are neglected and if, by using the tree level decay rate \(\Gamma(s)\), \(\text{Im}\Sigma(s)\) is only included at one loop. Since the process of the decaying \(\rho\) takes place in the centre of mass of the \(\tau\) lepton we require
\begin{align}
\frac{1}{N_\rho} \int\limits_0^{m_\tau^2} ds \ \! d_{\rho^- \rightarrow \pi^-\pi^0}(s) = 1 \ \!.
\end{align}
to ensure that the \(\rho\) fully decays in the \(\tau\) center of mass and that we do not include or leave out contributions that could come from higher-order perturbative corrections to the self-energy. 
We obtain the \(\tau\) spectral functions from
\begin{align}
d_W(s) = \frac{1}{\pi} \frac{\sqrt{s} \Gamma_{W}(s)}{M_w^4}\ \!,
\end{align}
where \(\Gamma_W(s)\) contains the coherent amplitude squared of the processes depicted in Fig. \ref{fig:FeynmanDiagramsVectorChannel} and we have used that for \(s \ll M_w^2\) the \(W\) propagator reduces to a point-like interaction vertex. 

\section{Results}
\begin{figure}[h]\begin{center}\includegraphics[width=0.77\textwidth]{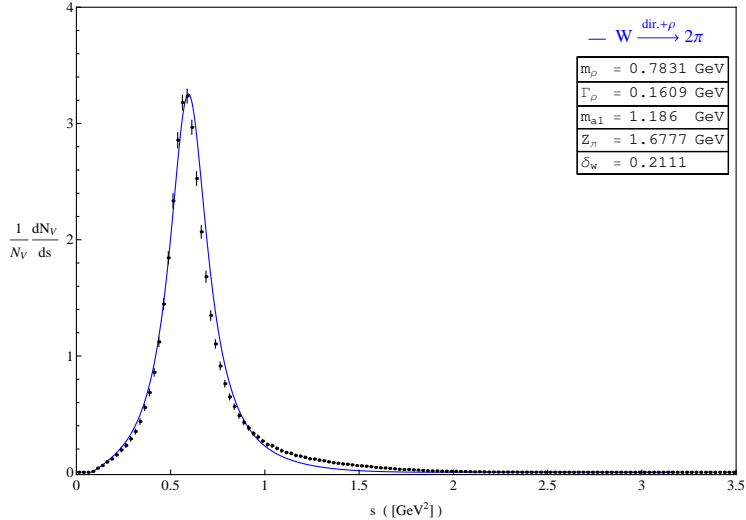}\end{center}
\vspace{-0.3cm}\caption{Spectral function in the vector channel of the coherent sum of the direct decay \(\tau^- \rightarrow W^- \nu_\tau \rightarrow \pi^-\pi^0 \nu_\tau\) and \(\tau^- \rightarrow W^- \nu_\tau \rightarrow \rho^- \nu_\tau \rightarrow \pi^-\pi^0 \nu_\tau\).}
\label{fig:01VectorChannelCoherent}\end{figure}

\begin{figure}[h]\begin{center}\includegraphics[width=0.78\textwidth]{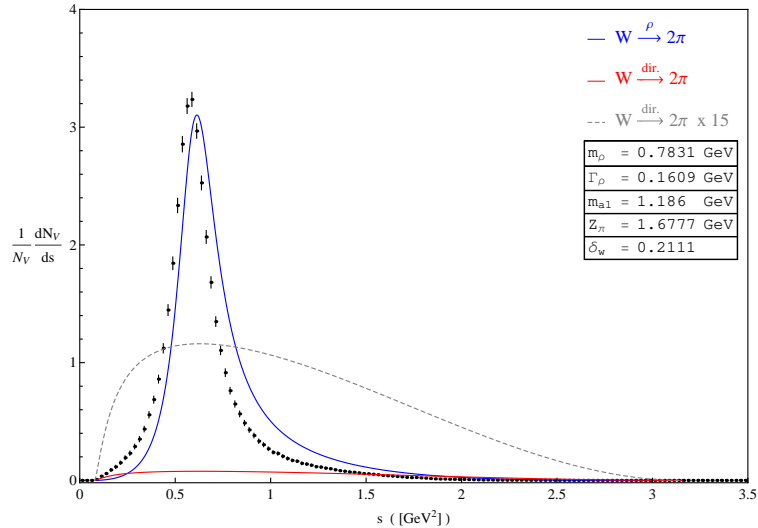}\vspace{-0.1cm}
\caption{The individual contributions to the spectral function. The blue line represents the contribution from \(W\!-\!\rho\) mixing, the red line the contribution from direct \(W^-\rightarrow \pi^- \pi^0\), and the dashed gray line the direct contribution scaled by \(15\).}
\label{fig:VectorChannelIndividualContributions}\end{center}\end{figure}

All relevant parameters, that is \(g_1\ \!,\ g_2\ \!,\ h_3 \ \!,\ w\), and \(Z\), except the mixing parameter \(\delta_w\), have been determined in a global fit in \cite{Parganlija:2012fy}. We use their result for masses and decay widths of \(a_1\) and \(\rho\) 
\begin{align}
&m_\rho = (0.7831 \pm 0.0070) \text{ GeV}\ \!,& \ &\Gamma_\rho=(0.1609 \pm 0.0044)\text{ GeV} \ \!, \\
&m_{a_1} = (1.186 \pm 0.006) \text{GeV}\ \!,& \ &\Gamma_{a_1} =(0.549 \pm 0.043 )\text{GeV}\ \! ,
\end{align}
such that \(\delta_w\) is the only free paramter. The result for the vector-channel spectral function compared to the ALEPH data is seen in Fig. \ref{fig:01VectorChannelCoherent}.  The coupling \(\delta_w\) that describes the \(W\!-\!\rho\) mixing is determined by the peak value of the mass distribution and is obtained as \(\delta_w = 0.2111\). Apart from a small shift of the peak to higher energies the spectral function nicely describes the experimental distribution \((1/N_{V}) dN_{V}/ds\). This shift is expected, since if we look at the results of the global fit in \cite{Parganlija:2012fy} we see that also there \(\Gamma_{\rho^- \rightarrow \pi^-\pi^0}\) deviates slightly from the experimental data. Within this fit all parameters have been determined to \(\pm 5\%\) accuracy. We can still adjust the parameters within this range and obtain an even better description. The individual contributions to the coherent sum are shown in Fig. \ref{fig:VectorChannelIndividualContributions}. It should be noted that \(\delta_w\) is determined from the coherent sum in Fig. \ref{fig:01VectorChannelCoherent}. If we consider only the contribution from \(W\!\!-\!\rho\) mixing we obtain a different height for the vector spectral function in Fig. \ref{fig:VectorChannelIndividualContributions} by choosing a different value for \(\delta_w\). The process \(\tau^- \rightarrow \pi^-\pi^0\) is clearly dominated by the intermediate \(\rho\) meson which constitutes almost the entire strength of the vector spectral function. However, the numerically small contribution of the direct \(W^- \rightarrow \pi^-\pi^0\) decay has a strong influence on the lineshape. Without the contribution from direct \(W^-\rightarrow \pi^-\pi^0\) the spectral function is shifted to higher values of \(s\) and there is an excess of the theoretical curve over the experimental data for \(s>m_\rho^2\). Including the direct \(W^-\rightarrow \pi^-\pi^0\) contribution into the coherent sum leads to destructive interference and removes the excess for \(s>m_\rho^2\). This can be interpreted as a very nice demonstration of VMD. In the low energy region the charged weak interaction fields seem to essentially couple to the pions by forming a vector meson resonance.

\section{Conclusion}
We have shown that the vector channel resonance in the electroweak decay of the \(\tau\) is very well described within the eL\(\sigma\)M together with a gauge invariant inclusion of an external electroweak field. Although from \(W\!-\!\rho\) mixing alone we already obtain a nice description of the \(\tau\) vector channel, we can clearly see that the modification of the spectral function by the direct contribution \(W^- \rightarrow \pi^-\pi^0\) is needed in order to reproduce the lineshape of the experimental data. We have also calculated the axial-vector channel of the \(\tau\) decay and the publication of this work is in progress.
\\~\\
{\bf Acknowledement:} The author acknowledges support from the Helmholtz Research School on Quark Matter
Studies and thanks D. H. Rischke and F. Giacosa for valuable discussions.

%uncomment the following lines to place a figure
%\begin{figure}[htb]
%\centerline{%
%\includegraphics[width=12.5cm]{Fig1}}
%\caption{Plot of ...}
%\label{Fig:F2H}
%\end{figure}

\end{document}